# Solution of mass/gap equations for strong velocity anisotropy in the $QED_3$ theory of underdoped cuprates

Dominic J. Lee

*Department of Physics, Simon Fraser University, Burnaby, British Columbia, Canada V5A 1S6*

(Dated: November 6, 2018)

The low-lying excitations at the nodes of the d-wave gap in the normal state for underdoped cuprates, close to the superconducting phase transition, may be described by an effective $QED_3$ theory. There are three characteristic velocities: $v_F$, $v_\Delta$ and $c$. For $v_\Delta = v_F = c$, the model reduces to $N$-flavour $QED_3$. Here, in the isotropic limit, for $N < N_c^0$, a critical number of fermions, a dynamical mass is generated which corresponds to the formation of a spin density wave. We study the effects of strong velocity anisotropy ($v_F \neq v_\Delta$) on dynamical mass generation. Solutions are given for the dynamical mass at both $N = 2$ and $N = 100$, and so we show that the critical number of fermions $N_c > 100$. However, we argue that $N_c = N_c^0$, when we go beyond the approximations used to derive our mass gap equations. We expect, though, that our solution for $N = 2$ is roughly correct, at low momentum, for large enough anisotropies. Implications and possible extensions to this work are also discussed.

## I. INTRODUCTION

Based on two simple assumptions, recently, it has been possible to derive an effective theory [1], [2] that describes the low energy excitations of the cuprates in the underdoped regime. Firstly, that there is gap of d-wave symmetry, to insure that there are gapless quasiparticle excitations at four points on the Fermi surface as well as a gap that changes sign. Experimentally, it has been reasonably well established that these materials possess d-wave symmetry. The results of [3], in particular, show a linear dependence in the penetration depth with respect to temperature, suggesting that there are, indeed, nodes in superconducting gap. In [4] there is further evidence to suggest that this assumption is correct; an observed change of sign with direction, in the superconducting gap, which is consistent with d-wave symmetry. In the underdoped regime, the cuprates can be considered as two dimensional [5]. Therefore, as a second assumption, one could reasonably expect that such materials undergo a Kosterlitz-Thouless type phase transition [6], in which vortex-antivortex pairs destroy the superconductivity. Already, there is some justification for this second assumption, through the experimental evidence of [7] and [8], that supports the idea of free vortices near and above the superconducting critical temperature.

The effective theory of [1], [2] is an anisotropic version of 2+1 dimensional quantum electrodynamics ($QED_3$), with three characteristic velocities; $v_F$ the Fermi velocity, $v_\Delta$ a velocity associated with the steepness of the gap at the 'nodes', and $c$ the propagation velocity of the gauge field. Here, the isotropic limit is realized when $v_F = v_\Delta = c$. The theory consists of *two* ($N = 2$) species of Dirac fermion that correspond to neutral spinons near the two pairs of diagonally opposed nodes in the d-wave state, as well as the gauge field that describes the interaction of the spinons with vortices. Spinons can be connected back to the real quasiparticle excitations in the cuprates through the singular gauge transformation discussed in [1],[2].

In [1] it was shown that at $T = 0$ there is inherent instability towards an antiferromagnetic state when one has unbound vortex loops in such a model. In $QED_3$ this instability corresponds to a "chiral" instability in which a mass is generated for fermions, dynamically, though their interaction with the gauge field. In the isotropic limit, such an instability is present when the number of Dirac components $N < N_c$, with $N_c \approx 3$ [9]. This number seems to be fairly robust, so as to stay roughly the same when more sophisticated approximations are employed [10],[11]. In the context of d-wave superconductivity, this dynamically generated mass is in fact a staggered potential felt by the original electrons, i.e. weak spin density-wave (SDW) order [1]. One should point out that such a spontaneously generated SDW automatically confines the neutral spin-1/2 excitations (spinons) of the superconductor [12], so ruling out spin charge separation when unbound vortices are present.

It is important to understand how velocity anisotropy in $QED_3$ affects the generation of a mass/gap. This problem is not only relevant to SDW generation in the cuprates, it also appears in the other effective Dirac theories describing condensed matter systems [13], which are not required to be Lorentz invariant. In [14] the effect of weak anisotropy was examined in the large-N approximation of [9], with two main results. Firstly, that weak anisotropy becomes *irrelevant* when the mass/gap, $\Sigma \to 0$; this was found to be in agreement with a similar study in 'chirally' symmetric state [15]. Secondly, it was argued that $N_c$ is the same in the weakly anisotropic case as it is in the isotropic case, if, indeed, $N_c$ can be considered universal. For the cuprates, however, the anisotropy is quite large $v_F/v_\Delta \sim 10$, [16]. So indeed one might ask what the effects of strong anisotropy might be.

We shall set out to answer such a question by first solving the gap equation in the strong anisotropic limit. Here, we employ the same approximation scheme as [14], but now we focus on the strong anisotropic limit. For our solution to the mass gap equation we find that $N_c > 100$. Secondly, we find for all our solutions that the mass/gap intially increases with momentum, this effect becomes more pronounced as the anisotropy is increased. We then go a step further and calculate the wavefunction renormalizations for each of the directions in momentum space. We find that $A_1 \approx 1$, where $A_1$ is the wavefunction renormalization in the direction most important for dynamical symmetry breaking in the strong anisotropic limit. This tells us that the approximation scheme we employ is likely to be reasonably accurate, provided that $N < N_c$, for strong anisotropies. We also calculate the renormalized anisotropy as a function of momentum, and discuss its effect on the validity of our approximation for strong anisotropy. Numerically, we find that $N_c > 100$, this may suggest either that $N_c$ is very large or infinite. If $N_c \to \infty$, we expect that this is an artifact the strong anisotropic limit, and that $N_c \to \infty$ only when the anisotropy tends to infinity. Furthermore, provided that $N_c$ is finite, we do not suspect that $N'_c$, the critical number of flavours calculated when the effects of renormalized anisotropy are considered self consistently in the Swinger-Dyson equations, is as large as $N_c$. Therefore, we would suggest that, for strong anisotropies, at some $N$ very close to $N'_c$ our approximation scheme breaks down. Based on two assumptions, $N_c$ being finite and universal, we argue that $N'_c = N_c^0$, the value calculated in the isotropic limit.

The layout of the paper is as follows. In Section II we derive the mass gap equation in the strong anisotropic limit and proceed to calculate an approximate solution using Pade approximants. In Section III, we derive equations for wavefunction renormalizations and the renormalized anisotropy in the strong anisotropic limit. Here, we calculate the effective values of the wavefunction renormalizations as well as the renormalized anisotropy. Finally, we present the argument why it could be the case $N_c = N_c^0$, its value in the isotropic limit. In Section IV, the disscussion, we discuss the limitations, and possible implications of our results.

## II. DYNAMICAL MASS GENERATION IN THE LIMIT OF STRONG ANISOTROPY

The effective low-energy theory for the d-wave quasi-particles (i. e. spinons) at $T = 0$ coupled to the topological defects (vortex loops [17] ) in the superconducting phase can be written as $\mathcal{L} = \mathcal{L}_1 + \mathcal{L}_2 + \mathcal{L}_a$ with [14] [1]:

$$\mathcal{L}_1 = i \sum_j^{N/2} \bar{\Psi}_{j,1} \left[ \gamma_0(\partial_0 - iea_0) + \frac{\delta}{\sqrt{\lambda}} \gamma_2(\partial_2 - iea_2) \right.$$

$$\left. + \delta\sqrt{\lambda} \gamma_1(\partial_2 - iea_1) \right] \Psi_{j,1}$$

$$\mathcal{L}_a = \frac{(\nabla \times \mathbf{a})^2}{2} + \frac{1}{2\epsilon} \mathcal{G}(\mathbf{a}) \quad (1)$$

and $\mathcal{L}_2$ is the same as $\mathcal{L}_1$, except that $1/\sqrt{\lambda} \leftrightarrow \sqrt{\lambda}$. Here, $\delta = \sqrt{v_F v_\Delta}/c$ and $\lambda = v_F/v_\Delta$. The $\gamma$-matrices satisfy the Clifford algebra $\{\gamma_\mu, \gamma_\nu\} = -\delta_{\mu\nu}$. The gauge field $\mathbf{a}$ represents the singular superconducting phase fluctuations induced by the vortex loops, and $\Psi$ fields represent the spinons; the neutral spin-1/2 excitations one can define near the nodes in the dSC. The coupling constant $e$ ("charge") is proportional to the dual order parameter that signals the appearance of infinitely large vortex loops, i.e. the destruction of the phase coherence in the dSC [1], [2]. We have assumed $N/2$ identical copies of each type of spinon field; the value of physical interest for a single layered superconductor being N=2. Here $\mathcal{G}(\mathbf{a})$ is a generalized gauge fixing term. We shall work in the landau gauge, as in [14] , so one chooses $\mathcal{G} = (\nabla . \mathbf{a})^2$ and $\epsilon = 0$. In the limit $\delta \to 1$, $\lambda \to 1$ the model simply reduces to the much studied $N$-flavour $QED_3$.

When studying dynamical mass generation in the isotropic limit, one may simply approximate the effect of Maxwell term simply by replacing $(\nabla \times \mathbf{a})^2$ by a u.v cutoff on the momentum integrals $\Lambda \approx Ne^2$. Essentially this can be done, because the effect of the Maxwell term is to cause $\Sigma$ to fall off rapidly at $p = \Lambda$, the momenta at which maxwell term becomes significant, so effectively cutting of the integral. But, considerable care must be taken in the strong anisotropic limit; the Maxwell term can no longer be neglected in such a way. We shall show that it is needed to regularize a i.r divergence present in the $k_0$ integral if one uses solely the polarization tensor in the strong anisotropic limit. Since, now, the Maxwell term can no longer be neglected it remains unclear what the role of $\delta$ is in the large $\lambda$ limit. This is because any non-trivial dependence on $\delta$ can no longer simply thought of as an artifact of gauge fixing [14], for scaling $\delta$ out of the action rescales both the gauge fixing term and the Maxwell term, which we are forced to retain. However, now for the sake of simplicity, in Eqn.1 we shall assume $\delta = 1$, and the problem of $\delta \neq 1$ will be left to a future publication.

Now, we shall calculate the mass gap for the type 1 spinons (those a appearing in $\mathcal{L}_1$) in the strong anisotropic limit. As discussed in [14], due to the $p_1 \to p_2$ symmetry of the theory, it is very easy to relate this to the mass gap for the type 2 spinons. From [14], we recall what the form of the full spinon propagator should take:

$$S_R(\vec{k}) = \left( \Sigma(\vec{k}) + A_\mu(\vec{k}) \gamma_\mu k_\mu \right)^{-1}. \quad (2)$$

Here, the repeated index, as usual, implies a sum over indices. $\Sigma$ is the dynamically generated mass-gap. $A_\mu$ is refered to as wave-function renormalization in $k_\mu$ direction. The starting point in our calculation of $\Sigma$ is the form of the gap equation given in [14] with $\delta = 1$

$$\Sigma(\vec{p}) = \frac{\alpha}{N} \int \frac{d^3\vec{k}}{(2\pi)^3} \frac{\Sigma(\vec{k})}{U(\vec{k})} \mathcal{K}(\vec{q}), \quad (3)$$

where

$$U(\vec{k}) = k_0^2 + \lambda k_1^2 + \frac{1}{\lambda} k_2^2 + \Sigma_1(\vec{k})^2,$$

$$\mathcal{K}(\vec{k}) = D_{0,0}(\vec{k}) + \frac{1}{\lambda}D_{2,2}(\vec{k}) + \lambda D_{1,1}(\vec{k}) \quad (4)$$

with $\vec{q} = \vec{k} - \vec{p}$, $\vec{k} = (k_0, k_1, k_2)$, and $\alpha = Ne^2$. Using this as well as the results given in [14] for the polarization tensor and the full gauge field propagator to one loop order, with massless spinons, one is able to extract the following limiting form for a gap equation in the Landau gauge.

$$\Sigma(\vec{p}) = \frac{16\sqrt{\lambda}}{N} \int \frac{d^3\vec{k}}{(2\pi)^3} \frac{\Sigma(\vec{k})}{\lambda k_1^2 + \Sigma(\vec{k})} \mathcal{H}(\vec{q}), \quad (5)$$

where

$$\mathcal{H}(\vec{q}) = \frac{q_1 q_2}{q^4} \left( \frac{16 q^2(q_0^2 + q_2^2)}{\alpha\sqrt{\lambda}} + q_0^2 \frac{(q_0^2 + q_2^2)}{|q_2|} \right.$$
$$+ |q|_2 (q_0^2 + q_2^2) + |q_1| q_2^2 ) \left( \frac{16^2 q_1 q_2 q^2}{\alpha^2 \lambda} + \right.$$
$$\left. \frac{16(|q_2|(q_0^2 + q_1^2) + |q_1|(q_0^2 + q_2^2))}{\alpha\sqrt{\lambda}} + q_0^2 \right)^{-1}. \quad (6)$$

It is not hard to see that if the terms arising from the maxwell term (terms multiplied by $1/\alpha\sqrt{\lambda}$ and $1/\alpha^2\lambda$ in Eqn.6) were neglected then one would, indeed, have a divergence in the $k_0$ integral at $p_0$. Next, we proceed by shifting the $k_0$ and $k_2$ integrals so that $H$ is independent of $p_0$ and $p_2$. Then, it is easy to show that there is a solution to Eqn.5 of the form $\Sigma(\vec{p}) = M(\tilde{p}_1)\alpha\lambda/16$ where $\tilde{p}_1 = p_1/(\alpha\sqrt{\lambda})$. On rescaling the momentum integrals we may write

$$M(\tilde{p}_1) = \frac{8}{\sqrt{\lambda}\pi^2 N} \int_0^\infty dk_1 \frac{M(k_1)}{k_1^2 + M(k_1)^2} \mathcal{U}(q_1), \quad (7)$$

where $\mathcal{U}(q_1) = \int dk_0 \int dk_2 \mathcal{H}(k_0, q_1, k_2)$, and now $q_1 = k_1 - \tilde{p}_1$. We first perform the $k_0$ integration; then when we come to performing the $k_2$ integration, we see that some of the terms contain logarithmic u.v divergences in $k_2$. To cure these divergences we first subtract out the singular part of $\mathcal{U}(q_1)$. We are able to show that the singular part of $\mathcal{U}(q_1)$ takes the following form

$$\mathcal{U}(q_1)_{sing} = G(q_1) \int_0^\infty dk_2 \frac{1}{1 + k_2}, \quad (8)$$

where

$$G(q_1) = \frac{0.375|q_1| + 0.25\sqrt{|q_1| + q_1^2}}{(\sqrt{|q_1|} + \sqrt{|q_1| + 1})^2} \quad (9)$$

These divergences should be considered an artifact of neglecting the $k_2^2$ term in $U(\vec{k})$. Therefore, to render our mass/gap equation completely finite we should then replace $\mathcal{U}(q_1)_{sing}(k_1^2 + M(k_1)^2)^{-1}$ with

$$\mathcal{U}(q_1)_{cor} = G(q_1) \int_0^\infty dk_2 \frac{1}{1 + k_2} \frac{1}{k_1^2 + k_2^2/\lambda^2 + M(k_1)}. \quad (10)$$

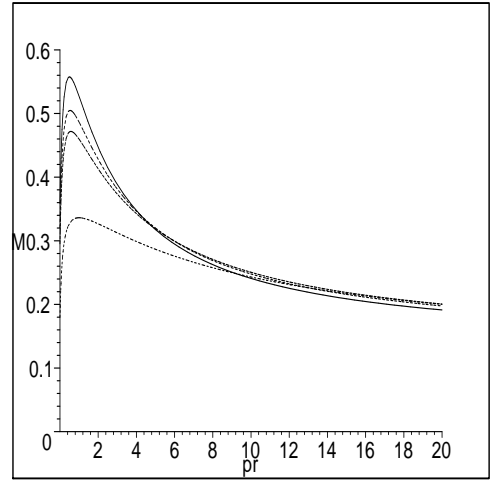

FIG. 1: Graph showing $M$ calculated for $N = 2$, $\lambda = 10, 15, 20, 100$, as a function of $\tilde{p}_1$, the rescaled momenta. The curve with the smallest magnitude is for $\lambda = 100$ and the largest curve is for $\lambda = 10$, the curves decreasing in size with $\lambda$

On evaluating this integral for large $\lambda$ we are then able to recast our gap equation in a final form

$$M(\tilde{p}_1) = \frac{8}{\sqrt{\lambda}\pi^2 N} \int_0^\infty dk_1 \frac{M(k_1)}{k_1^2 + M(k_1)^2} \mathcal{F}(k_1, q_1),$$
$$\mathcal{F}(k_1, q_1) = [\mathcal{U}(q_1)_{fin} + \log(\lambda)G(q_1) + G(q_1)$$
$$\times \left( \frac{\pi}{(k_1^2 + M(k_1)^2)^{1/2}} - \log(k_1^2 + M(k_1)^2) \right) \right], \quad (11)$$

where $\mathcal{U}(q_1)_{fin} = \mathcal{U}(q_1) - \mathcal{U}(q_1)_{sing}$. The $k_2$ integral for $\mathcal{U}(q_1)_{fin}$ needs to be evaluated numerically, we find that $\mathcal{U}(q_1)_{fin}$ is reasonably well approximated by

$$\mathcal{U}(q)_{fin} \approx \frac{(.01357 + 62.84|q| - 39.04q^2 - 5.476|q|^3)}{(.3109 + 88.55|q| + 78.61q^2 + 3.486|q|^3)}. \quad (12)$$

As a further approximation we also replace $M(k_1)$ with $M(0)$ in all the dominators as well as the logarithm, an approximation valid for $M < 1$. Then, we calculate an approximate solution for the mass gap equation using an approximate solution of the form

$$M(p) = \frac{m_1 + m_2|p| + m_3 p^2}{1 + m_4|p| + m_5 p^2}, \quad (13)$$

then by calculating $M(0), M(1/3), M(2/3), M(1)$ and $M(5)$ for trial values of $m_i$ ($i = 1..5$) one calculates new values of $m_i$. We then iterate the process until the $M(p)$ inserted into R.H.S is the very close to the $M(p)$ calculated for L.H.S in Eqn.11

In Fig.1 we show the calculated values of the mass/gap for $N = 2$ for $\lambda = 10, 15, 20, 100$. As $\lambda$ is increased the solutions are seen to fall in magnitude. One can see that all the solutions seem to possess three qualitative features. The first is that for small momenta we see that the mass/gap increases with rescaled momenta. Then a maximum in $M(p_1)$ is reached and then the solutions decrease with momenta. Finally, we see that the solutions tend off to a constant for large momenta. However, as we discuss later in section IV, this last feature may well be an artifact of our approximation. As $\lambda$ is increased the first and third feature become more pronounced, whereas the solutions seem to drop off less for momenta above the maximum in $M$.

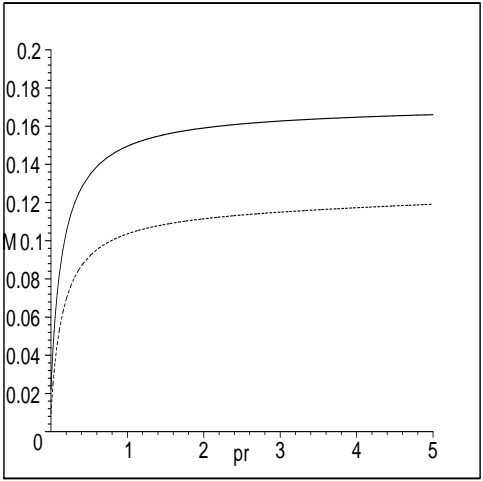

FIG. 2: Graph showing $M$ calculated for $N = 100$, $\lambda = 10, 100$, as a function of $\tilde{p}_1$, the rescaled momenta. The curve with the smaller magnitude is for $\lambda = 100$ and the larger curve is for $\lambda = 10$.

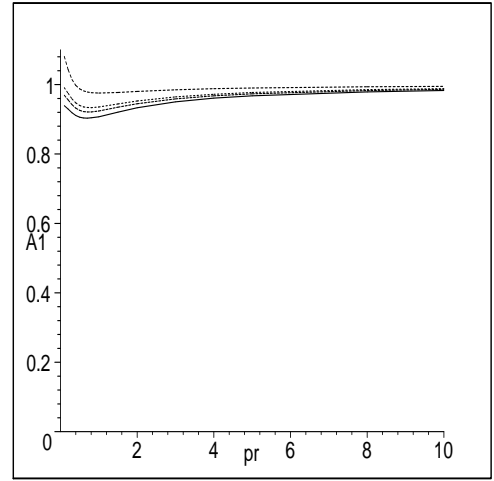

FIG. 3: Graph showing $A_1$ calculated for $N = 2$, $\lambda = 10, 15, 20, 100$ in units of $\sqrt{\lambda}$, as a function of $\tilde{p}_1$, the rescaled momenta. The smallest curve is for $\lambda = 10$ and the largest curve is for $\lambda = 100$, the curves increasing in size with $\lambda$

In Fig.2 we show the calculated values of the mass/gap for $N = 100$ for $\lambda = 10, 100$. Again, we see that as $\lambda$ is increased the solutions are seen to fall in magnitude. However, we see that no-longer a maximum in $M$ can be seen; these solutions seem to be monotonically increasing with momentum towards some constant value. Since a solution exists for $N = 100$ we may conclude that $N_c > 100$ for Eqn. 11, where only bare anisotropies are considered.

## III. THE WAVEFUNCTION RENORMALIZATIONS AND RENORMALIZED ANISOTROPY

From [14] we have the following expressions for the $A_\mu$

$$A_\mu(\vec{p}) = \lambda_\mu \left(1 - \frac{\alpha}{N p_i} \int \frac{d^3 \vec{k}}{(2\pi)^3} \frac{2\mathcal{T}_\mu(\vec{k}, \vec{q}) - k_\mu \mathcal{K}(\vec{q})}{U(\vec{k})}\right) \quad (14)$$

where

$$\mathcal{T}_\mu(\vec{k}, \vec{q}) = k_0 D_{\mu,0}(\vec{q}) + \lambda k_1 D_{\mu,1}(\vec{q}) + \frac{k_2 D_{\mu,2}(\vec{q})}{\lambda}. \quad (15)$$

Here, $\lambda_0 = 1$, $\lambda_1 = \sqrt{\lambda}$ and $\lambda_2 = 1/\sqrt{\lambda}$. From these expressions one is able to show in the large $\lambda$ limit that the $A_\mu(\vec{p})$ depend only on $p_1$ and

$$A_1(p_1) = \sqrt{\lambda} \left(1 - \frac{16\sqrt{\lambda}}{N p_1} \int \frac{d^3 \vec{k}}{(2\pi)^3} \frac{k_1 \mathcal{H}(\vec{q})}{\lambda k_1^2 + \Sigma(k_1)^2}\right),$$
$$A_0(p_1) = \sqrt{\lambda} A_2(p_1) = 1 - \frac{16\sqrt{\lambda}}{N} \int \frac{d^3 \vec{k}}{(2\pi)^3} \frac{\mathcal{H}(\vec{q})}{\lambda k_1^2 + \Sigma(k_1)^2}, \quad (16)$$

where $\Sigma(k_1) = \Sigma(0, k_1, 0)$. Then it is easy to see, that to calculate the $A_\mu$, one goes through the same analysis discussed in the previous section. We arrive at the following expressions from Eqn.16

$$A_0(\tilde{p}_1) = 1 - \frac{8}{\sqrt{\lambda}\pi^2 N} \int_0^\infty dk_1 \frac{\mathcal{F}(k_1, q_1)}{k_1^2 + M(k_1)^2},$$
$$\frac{A_1(\tilde{p}_1)}{\sqrt{\lambda}} = 1 - \frac{8}{\sqrt{\lambda}\pi^2 N p_1} \int_{-\infty}^\infty dk_1 \frac{k_1 \mathcal{F}(k_1, q_1)}{k_1^2 + M(k_1)^2}. \quad (17)$$

In Fig.3 we calculate $A_1$ for $N = 2$, $\lambda = 10, 15, 20, 100$; also the renormalized anisotropy $\lambda_r = \sqrt{\lambda} A_1(p_1)/A_0(p_1)$, again for $N = 2$, $\lambda = 10, 15, 20, 100$ in Fig.4. In examining Fig.3, the first thing to notice is that $A_1$ does not depend much on $p_1$ and is close to 1. This means that the effect of including vertex corrections, as well as wave function renormalization for the fermion propagator, in the self energy might be very small. This is assuming the vertex ansatz $\Gamma_\mu = A_\mu \gamma_\mu$ (or some other ansatz depending only on $A_\mu$) is a good approximation, where $\Gamma_\mu$ is the full vertex function. This ansatz is simply a generalization of the ansatz $\Gamma_\mu = A\gamma_\mu$ used in the literature when considering the isotropic limit, [11] [18]. This assumes that $\Sigma$ can be neglected in vertex corrections.

In Fig.4 we see that the renormalized anisotropy is effectively a half of its bare value, and changes little with momenta. Although this does effect the equation for the self energy, which depends only on $A_1$ and $\Sigma$ in strong anisotropic limit, this may well effect the polarization tensor, though by how much is not clear. This is simply due to the fact that all elements of the polarization tensor, with the exception of $\Pi_{1,2}$, contribute in the strong anisotropic limit. This implies that all the $A_\mu$ contribute. However, such an effect on the polarization may lead to a small change in $\Sigma$. More importantly, $\lambda_r$ will effect at what values of $\lambda$ our large momentum approximation is valid. The strong anisotropic limit should be considered to be a good approximation above some specified value $\lambda = \lambda_s$, so far undetermined. The fact that the renormalized anisotropy is approximately half the bare value tells us that $\lambda_s$ should be effectively doubled, when the renormalized anisotropy is included in the self energy.

Now we present an argument that $N_c(\lambda) = N_c(\lambda = 1)$. There are two assumptions that we shall make. The first is that $N_c(\lambda)$ is universal; that $N_c(\lambda)$ depends only on $\lambda_r(0)$ in the limit $M(0) \to 0$. The second is that for finite $\lambda$ there exists a 'bare' (calculated with only the bare anisotropy $\lambda$) finite $N_c$, which monotonically increases so that $N_c \to \infty$ as $\lambda \to \infty$. Such a finite $N_c$ probably arises from terms neglected in the strong anisotropic limit, as our result $N_c > 100$ suggests that $N_c \to \infty$ for Eqn.11. At such an $N_c$, $M(0) = 0$. If we are able show that for $M(0) \to 0$ that $\lambda_r$ flows towards zero, then, the effect of a falling $\lambda$ is to reduce $N_c$. We know from [14], [15] that $\lambda_r(0)$ flows towards 1 for weak anisotropies, so if it does so for strong anisotropies then it is probably safe to assume that $\lambda$ is irrelevant for all values. So, assuming that these assumptions are correct, then indeed it must be the case that $N_c(\lambda) = N_c(\lambda = 1)$. We are able show that $\lambda_r(0)$ is irrelevant in the strong anisotropic limit.

To do this we must examine both expressions in Eqn.17

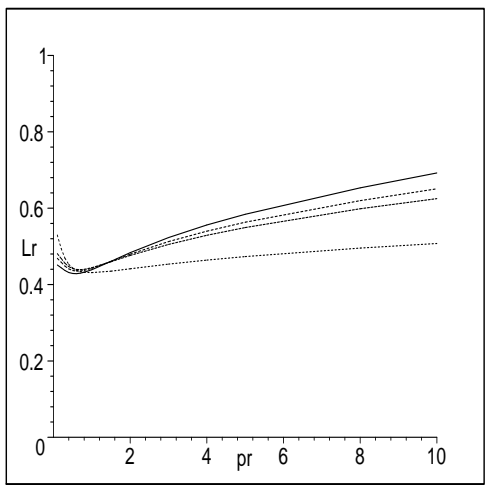

FIG. 4: Graph showing $\lambda_r$ calculated for $N = 2$, $\lambda = 10, 15, 20, 100$, in units of $\lambda$, as a function of $\tilde{p}_1$, the rescaled momenta. The curve with the smallest magnitude for large rescaled momentum is for $\lambda = 100$ and the largest curve is for $\lambda = 10$, the curves decreasing in size with $\lambda$ for large momenta

in the limit $M(0) \to 0$, for $p_1 = 0$. In the limit $p_1 \to 0$ we notice that

$$\frac{A_1(\tilde{p}_1)}{\sqrt{\lambda}} = 1 + \frac{8}{\sqrt{\lambda}\pi^2 N} \int_{-\infty}^{\infty} dk_1 \frac{k_1 \left.\frac{\partial \mathcal{F}(k_1,q_1)}{\partial q_1}\right|_{p_1=0}}{k_1^2 + M(k_1)^2}. \quad (18)$$

Because of the singular i-r behaviour of both expressions in the limit $M(0) \to 0$, we require only the small $q_1$ behaviour of both $\mathcal{F}(k_1, q_1)$ and its derivative with respect to $q_1$. By extracting the small $q_1$ behaviour of $\mathcal{F}(k_1, q_1)$ we are able to show that the dominant singular behaviour in the limit $M(0) \to 0$ is

$$A_1 \approx \sqrt{\lambda}\left(1 + \frac{2}{\pi^2\sqrt{\lambda}M(0)^{3/2}}\right),$$
$$A_2 \approx \frac{1}{\sqrt{\lambda}}\left(1 + \frac{4}{\pi^2\sqrt{\lambda}M(0)^{3/2}}.\right) \quad (19)$$

From these two expressions we may assume for $\lambda >> 1$

$$\lambda'_r \approx \lambda' - \frac{2\sqrt{\lambda'_r}}{\pi^2 M(0)^{3/2}} \quad (20)$$

where $\lambda' = \lambda - 1$ and $\lambda'_r = \lambda_r(0) - 1$ (this is certainly true to $O(1/N)$). Then Eqn.20 suggests that $\lambda'_r \sim M(0)^3$, so we conclude $\lambda'_r$ decreases towards zero.

## IV. DISCUSSION

In Section II we calculated the mass gap equation for $N = 2$ and $\lambda = 10, 15, 20, 100$, as well as $N = 100$ and $\lambda = 10, 100$. An important feature, pointed out in Section III for these solutions, is that for large momentum $m(p_1)$ tends to a constant. This feature is likely to be a artifact of our approximation [19], and in reality we might expect our solutions to fall off slowly at large rescaled momenta $\tilde{p}_1$ for large, but finite $\lambda$. However, we expect that our solutions are reasonable approximations at small rescaled momenta. What is reassuring in regards to this conclusion, is that we find that the wave function renormalization $A_1$ calculated in Section III is close to one. As discussed there, this we would take as a sign that the approximation scheme we employ to derive Eqn.5, our intial mass/gap equation, is reasonably accurate in the regime of strong anisotropy. So our conclusion would be that the feature of the mass gap intially increasing with momentum $p_1$ should certainly be true for $QED_3$ for large $\lambda$.

We should point out that the situation could be completely different in the cuprates. Since the mass/gap is no longer cutoff at large momenta by the Maxwell term in our approximation, one would expect that some higher order term like the operator $(\nabla^2 \times \mathbf{a})^2/e^2$ would cut the integral off at some scale $\tilde{p}_1 \sim 1/(N\lambda)$, thereby leading to significantly different behaviour. Such terms are present in the full effective theory [20],[21] for the cuprates, once the vortex order parameter $\Phi$ and auxiliary gauge field, through which it couples to **a**, have been integrated out in the non-superconducting phase.

Our most significant result seems to be that we can provide evidence for $N_c(\lambda) = N_c^0$, provided that our assumptions are correct. Namely, that $N_c$ is universal and there exists a finite $N_c$, at finite $\lambda$ for $\Sigma$ calculated with only the bare anisotropy in the mass gap equation. Even if the first assumption that we make is not correct for large anisotropies, we should take the fact that $\lambda_r$ decreases as $M(0) \to 0$, as an indication that $N_c(\lambda)$ changes little with $\lambda$. The result that $N_c > 100$ suggests that it could be that $N_c \to \infty$, but we should take this as $N_c$ for the limiting case where $\lambda \to \infty$. Detecting a finite $N_c$, or not, for bare anisotropy is likely to go beyond our strong anisotropic approximation, so requiring a more careful analysis of Eqn.5. However, it would seem strange that the second assumption, a finite $N_c$, did not hold. This would imply a $\lambda_c$ above which $N_c \to \infty$. We think this is unlikely. The argument is unaffected by the inclusion of higher order operators, for example $(\nabla^2 \times \mathbf{a})^2/e^2$, as it relies on the form of the small $q_1$ dependence of $\mathcal{F}(k_1, q_1)$, where such terms can be safely neglected. The value that $N_c$ takes for strong anisotropies has profound implications in the model discussed in [21]. If $N_c$ was to exceed 10 for anisotropies where $\lambda \sim 10$, might it be possible in some way to have coexistence of both antiferromagnetism and superconductivity. If our assumptions hold then the argument given in [21], that there one phase transition, must hold for strong anisotropies. The conclusions, here again, suggest that for a single layered cuprate ($N = 2$) that there is no intermediate phase [22], [23] between the SDW and the superconducting phase at $T = 0$.

Clearly, an interesting problem for future consideration is the effect of higher order operators in the gauge field of $O(k^4)$ or higher on the solution of the mass gap equation. Even, in the isotropic limit the quantitative effect of these terms might be significant, even though such terms are unlikely to change qualitative behaviour of the mass/gap. For instance, such terms might affect the value of the momentum at which the mass/gap is effectively cutoff at $N = 2$, but such a value will still be proportional to $e^2$. As discussed before, the effect of such terms for large anisotropy is likely to become more pronounced, possibly leading to major qualitative differences. Another interesting problem is what effect $\delta \neq 1$ might have on the solutions to the mass gap equation for large $\lambda$, this again is a problem we would like to address. There also remains the tricky issues of the universality of $N_c$ and the form of the exact solution of Eqn.3, valid for all $\lambda$, but these are much harder problems.

## V. ACKNOWLEDGEMENT

This work is supported by NSERC of Canada and the Research Corporation. D.J. Lee would also like to thank Igor Herbut for useful discussions and critical reading of the manuscript.